\documentclass{emulateapj}
\usepackage{times}
\usepackage{graphicx}

\slugcomment{To Appear in ApJ}

\shorttitle{The $\gamma$-ray Pulsar Population}

\begin{document}

\title{The Galactic Population of Young ${\gamma}$-Ray Pulsars}

\author{Kyle P. Watters\altaffilmark{1} \& Roger W. Romani\altaffilmark{1}}
\altaffiltext{1}{Department of Physics, Stanford University, Stanford, CA 94305}
\email{kwatters@stanford.edu, rwr@astro.stanford.edu}

\begin{abstract}
	
We have simulated a Galactic population of young pulsars
and compared with the {\it Fermi} LAT sample, constraining the birth
properties, beaming and evolution of these spin-powered objects.
Using quantitative tests of agreement with the distributions of observed spin and pulse
properties, we find that short birth periods $P_0 \approx 50\,$ms and 
$\gamma$-ray beams arising in the outer magnetosphere, dominated by a 
single pole, are strongly preferred. 
The modeled relative numbers of radio-detected and radio-quiet objects 
agrees well with the data. Although the sample is local, extrapolation
to the full Galaxy implies a $\gamma$-ray pulsar birthrate 1/(59 yr).
This is shown to be in good agreement with the estimated Galactic 
core collapse rate and with the local density of OB star progenitors.
We give predictions for the numbers of expected young pulsar detections
if {\it Fermi} LAT observations continue 10 years. In contrast to
the potentially significant contribution of unresolved millisecond
pulsars, we find that young pulsars should contribute little to 
the Galactic $\gamma$-ray background.
\end{abstract}


\section{Introduction}

        The {\it Fermi Gamma-ray Space Telescope} has proved remarkably 
successful at discovering spin-powered pulsars emitting GeV $\gamma$-rays
\citep{psrcat}. The brighter sources can be studied in great detail and
in our quest to understand this emission we have shown how careful 
multiwavelength analysis and geometrical modeling of individual sources can
improve our understanding of the $\gamma$-ray beaming \citep{wat09,rw10}.
In addition, the {\it Fermi} LAT sensitivity has allowed many
new pulsar detections; the reported $\gamma$-ray pulsar numbers
have increased by more than an order of magnitude since launch in 
June of 2008. This allows, for the first time, a serious study
of this pulsar population. Such work provides further guidance to
the correct beaming models, and also constrains the birth and evolution
of these energetic pulsars, with important implications for the
study of Galactic supernovae and their products.

	Here we focus on the young $\gamma$-ray pulsars, whose 
connection with massive stars is particularly important. These objects,
with $\sim 0.03-0.3$\,s periods, have magnetospheres extending to
many $R_{NS}$ so that dipole fields should dominate away
from the surface, allowing relatively simple models. {\it Fermi}
is also discovering a large number of recycled millisecond pulsars (MSP).
We do not treat these objects here, since their very
compact magnetospheres may allow departures from dipole field
geometries to complicate the magnetosphere modeling; additionally their
long lives and rare binary formation channels make population synthesis
more challenging. Early attempts to describe the population of
$\gamma$-ray MSP \citep{story08,fgl10} suggest the importance of this
source class. However the recent spate of {\it Fermi} detections
requires a re-assessment of these models which we defer to future
analysis.

	We start by describing our simulation method and the process
used to build a model galaxy of pulsars (\S \ref{Simulation}).
We next describe the different $\gamma$-ray models to be
applied to this simulated galaxy and study the distribution of
$\gamma$-ray pulse profile morphologies that result (\S \ref{PulseMorph}).
These are compared with the observed pulse profiles from {\it Fermi}.
We then discuss the modeled evolution of $\gamma$-ray emission with age and 
decreasing spin-down luminosity,
especially near the $\gamma$-ray death line (\S \ref{Evolution}). Some
possible amendments to the $\gamma$-ray model and their
observational effects are discussed (\S \ref{Amendments}).
We conclude in \S \ref{Conclusion} by describing the unresolved young
pulsar background and predicting the properties of pulsars accessible to
{\it Fermi} with ten years of observations.

\section{Simulating the Galactic Young Pulsar Population}
\label{Simulation}

	There is a long history of pulsar population synthesis studies
using very detailed modeling of the radio emission properties and
the survey selection effects so that comparison with the large
radio samples can be used to constrain the population. The recent effort
of \citet{fg06}, in particular, has been quite sucessful at reproducing
the properties of the bulk of the radio pulsar population. Of course,
there have been earlier efforts to extrapolate from the radio pulsars
and estimate the numbers of $\gamma$-ray
detectable pulsars \citep{ry95, get07}. However, with the large increase
in pulsar numbers and the improved understanding of $\gamma$-ray beaming 
supplied by the LAT, it is now possible to use this sample to produce 
much more reliable modeling of the young pulsars.
This is the goal of the present paper.

	Excluding the recycled pulsars, the LAT sample is very energetic
($\dot E \gtrsim 5\times10^{33}\, \rm erg\,s^{-1}$)
and young ($\tau \lesssim 10^6\,$yr). It is, moreover, a more complete sample of the
nearby energetic pulsars than available from the radio alone. 
For example the ATNF pulsar catalog contains three pulsars with spin-down
luminosity ${\dot E} > 10^{35}\,{\rm erg \, s^{-1}}$ and $d \le 2\,{\rm kpc}$.
The LAT young pulsar sample has at least six such objects and four additional
pulsars with uncertain distances having a large overlap with this region.
Thus, study of the $\gamma$-ray pulsars provides an independent, and
arguably better, picture of the birth properties of energetic pulsars.

	Such study requires a realistic model of the birth locations, 
kinematics, and spin properties of energetic neutron stars. Much of our
treatment is quite standard. For example, we
assume the birth velocity distribution of \citet{ho05}
and assign each modeled pulsar
a birth velocity with random isotropic direction and 
magnitude drawn from a three-dimensional Maxwellian
with $\sigma = 265\, \rm km\,s^{-1}$ in each dimension.
The pulsar is then shifted from its birth position to the 3D
location expected at the appropriate simulated age. We safely
ignore the Galactic acceleration's effect on these trajectories for
this young $\tau\lesssim 10^6\,$yr sample.
We use a conventional birth magnetic field distribution,
log-normal with $\langle \rm{log}(B)\rangle =12.65$, $\sigma_{\rm{log}(B)} = 0.3$, 
with B in Gauss, (similar to the values used in \citealt{fg06} and \citealt{yr95}).
We also assume isotropic distributions
of magnetic inclination angle $\alpha$ and Earth viewing angle $\zeta$,
and a conventional width for the radio beam $\rho$. Indeed, these choices
seem quite robust as even modest changes lead to unacceptable populations. 
We select the true age uniformly up to $10^7\,$yr (to ensure that we cover
the full range over which pulsars may possibly still be active in the $\gamma$
rays) and evolve the pulsars at
constant $v$ and $B$ to the present, where they are observed at their 
appropriate evolved spin-down luminosity ${\dot E}$. Because we focus
here on only the most energetic pulsars, we can adopt such simple 
constant evolution; high spin-down energy corresponds to low age, and the
majority of our simulated $\gamma$-ray active pulsars have $\tau \lesssim 10^6$ yr.
However, this means that we do not attempt to model
the old radio pulsar sample, where the effects of gravitational acceleration,
field decay and the approach to the radio death line can affect detection
statistics. We do not simulate here the binary star properties.
We also do not attempt to follow the details of the individual
radio surveys whose samples are dominated by these older objects. However, we 
do apply a uniform sensitivity threshold ($\rm{F_{min}}=0.1\,\rm{mJy}$) and 
assume that pulsars are radio-detected if $|\zeta - \alpha| < \rho$ and
the modeled radio flux exceeds this sensitivity at the modeled distance from Earth.
This is adequate to compare the radio and $\gamma$-ray detectabilities and 
check the overall population normalization against the radio surveys,
with special attention to the large and uniform Parkes Multibeam Survey (PMBS) radio 
sample \citep{man01,lor06}.

	Our focus on the young pulsars and our interest in connecting to
their massive star progenitors motivates some amendments to the standard
modeling.  In the next section we develop a detailed model of the progenitor 
distribution to allow us to connect our pulsar numbers to the OB stars and
our local estimates to the Galaxy as a whole. We also find (in \S \ref{RadioSpin})
that the LAT sample requires amendments to the treatment of the
birth spin and the radio luminosity of the pulsars. Finally, we study
in detail the effect of the evolving $\gamma$-ray beams on the detection numbers
and pulse properties. These extensions give a new view of the energetic pulsar
sample at birth, with important implications for the neutron star
population as a whole.

\subsection{Galactic Structure}
\label{GalStruct}

The non-recycled $\gamma$-ray pulsars almost exclusively have ages $< 10^6$\,yr, 
substantially less than those of their parents, dominantly
$\sim 10M_\odot$ B stars. Thus it is not unexpected that
their distribution along the Galactic plane correlates with 
young star-forming regions \citep{yr97}. To exploit this correlation
we need a detailed model of Galactic birth-sites, namely high mass
OB stars. On the largest scale we follow \citet{yr97} in using the
free electron distribution (here we use the updated model of \citealt{cl02}) 
as a proxy for the gas
and ionizing photons associated with massive stars. This model includes
the Galactic spiral arms, the 3.5\,kpc ring and an exponential thin disk 
component with a scale height of 75 pc, which we take here as the OB star
plane distribution. High mass stars also show a ``runaway'' component,
from disrupted binaries, giving roughly 10\% of the stars peculiar 
velocities greater than $30\ {\rm km\ s}^{-1}$ \citep{dray06}. We account
for this component with a exponential thick disk, as in the
Cordes \& Lazio model, but with a scale height of 500 pc.

\begin{figure}[h!]
\vskip -0.8truecm
\hskip -0.7truecm
\includegraphics[scale=0.47]{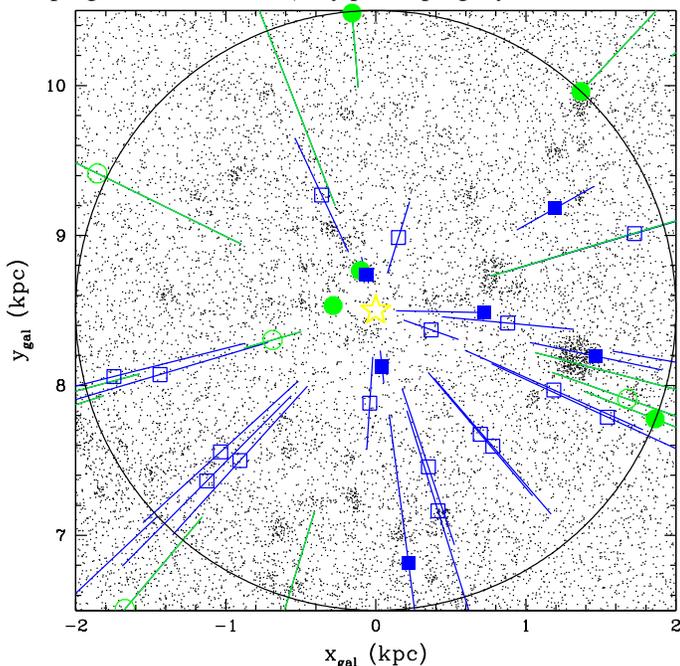}
\vskip -0.4truecm
\caption{\label{LocalGalaxy} Modeled OB stars (i.e. pulsar birth-sites)
in the nearby Galactic plane (black dots). Detected $\gamma$-ray
pulsars are also indicated (radio-selected as green circles,
$\gamma$-selected as blue squares, for the distinction between
these two classes, see their definitions in \S \ref{Samples}).
Radial lines indicate the uncertainty in the distance estimates
(open symbols indicate pseudo- and
DM- distance estimates, filled symbols indicate parallax, kinematic, and 
association-based distances). The Galactic Center lies at (0, 0).
The overdensity of birth sites
at Galactic radii just larger than the Suns' 8.5\, kpc represents
the Orion spur. Cyg OB2 is prominent toward the right side. 
Inner arm clusters appear toward the bottom of the plotted region.}
\end{figure}

	The LAT pulsar population is apparently quite local 
(mostly $\lesssim 2$\,kpc) and on these scales we can employ more detailed information
on the parent star distribution. In particular, the Hipparcos catalog
gives a nearly complete sample of O and B stars to $\sim 500\,$pc
with useful parallax information. On somewhat larger scales, catalogs
of OB associations \citep{mel95} give the size and locations of massive 
star concentrations. We have used these data to generate a 3-D model of
the Galactic O and B0-B2 star density. Locally, we use the direct 
Hipparcos sample of such stars with a $\ge 2\sigma$ measurement of parallax
(smoothed by a 50\,pc spherical Gaussian). At 375\,pc we make a smooth
transition to a sample with uniform disk distributions (thin + thick runaway)
augmented by the cluster OB star contribution (smoothed over 100\,pc).
In turn, this population merges smoothly to the large scale (spiral arm
plus disk) distribution with a transition at 1.7\,kpc.
The overlap between the individual stars and clusters and between the
clusters and the spiral arms allows us to match all components to
the normalization determined by the local, complete OB star sample.
The massive stars (i.e. pulsar birth-sites) drawn from this distribution
are shown in  Figure \ref{LocalGalaxy} projected to the nearby Galactic 
plane.  Although the uniform component is substantial, large OB 
concentrations and clusters are apparent.

	Figure \ref{LocalGalaxy} also shows the projected location 
of the nearby LAT pulsar sample. For many of these objects the 
distance estimates are quite poor (range shown by radial lines). Thus while several
pulsars are superimposed on massive star associations,
the poor distance constraints prevent confident assignment.
We can, however, form a `Figure of Merit' to test agreement
between the distribution of observed pulsar positions and
positions from our detailed model simulation.
This is an overlap sum over the set $\{i\}$ of model birthsites with a weight
determined from a Gaussian distribution about the uncertain location of
each observed pulsar, $j$. This distribution combines the (generally large) 
uncertainty in the pulsar radial distances with a transverse Gaussian spread of 50 pc to 
account for the smoothed birthsite distribution and the offset
due to pulsar proper motion. The result is
\begin{equation}
{\rm FoM} = \sum_{j}\,\sum_{i}e^{-\frac{(l_j-l_i)^2+(b_j-b_i)^2}{2\tan^{-1}(50\,{\rm pc}/d_j)^2}} \times e^{\frac{-(d_j-d_i)^2}{2\sigma_{d_j}^2}}.
\end{equation}
Errors were estimated by boostrap analysis.

	Comparisons were made between this smoothed simulation and
the LAT pulsars. The detailed model gives a value of $156\pm33$. This is
slightly, but not significantly, better than the values found from 
pulsars distributed according to the \citet{cl02} spiral arm model 
($136\pm22$) or a simple uniform exponential disk alone ($147\pm26$).
We will thus use our detailed model in further analysis,
although we caution that it is not yet demanded by the data. 
Improved pulsar distance estimates and/or analysis that uses observed
pulsar proper motions to correct back to birth-sites can
substantially improve the utility of the detailed Galactic
OB distribution.

	One relatively robust result from the OB star modeling is
a connection with Galactic core collapse rates. Summing up the
local, nearly complete, Hipparcos sample for each luminosity class
and using each class' nuclear evolution lifetime \citep{reed05}, 
we can calculate the supernova rate contribution from each. 
We find, including stars from B2-O5, a local supernova rate 
of $40\,{\rm SN\,kpc^{-2}\ Myr}^{-1}$ (very high mass stars
which might produce black holes contribute $\sim 1\%$ of this rate).
In turn our model allows us to extrapolate this to the Galaxy as a whole,
yielding a Galactic O-B2 star birth rate (i.e. SN rate) of 
$2.4/ {\rm 100\,yr}^{-1}$.  This compares very well with other recent estimates, eg.
$2.30\pm0.48/ {\rm 100\,yr}^{-1}$ from nearby supernovae \citep{li10}, 
$1.9\pm1.1/ {\rm 100\,yr}^{-1}$ from $^{26}$Al emission \citep{die06}
and $\sim1-2 / {\rm 100\,yr}^{-1}$ from a similar analysis of 
nearby massive stars \citep{reed05}. Thus our model is well normalized
and can be used to connect O-B2 progenitors with their $\gamma$-ray 
pulsar progeny.

\subsection{Birth Spin Distribution and Pulsar Radio Emission}
\label{RadioSpin}

	The new-born pulsar population depends critically on
the distribution of spin periods at birth. This is especially important
for understanding the $\gamma$-ray sample; the bulk of the radio population
is sufficiently spun down to retain little memory of this adolescent
phase.  We describe the birth spin periods with a single parameter truncated
normal distribution:
\begin{equation}
{\rm Prob} (P)\propto e^{-\frac{(P-P_0)^2}{P_0^2}}.
\end{equation}
Here the spread equals the mean and we truncate at a minimum $P=10$\,ms.

	\citet{fg06} find a characteristic birth spin period of
$P_0 = 300$\,ms for the radio population. However, we find that
this produces very few short period, $\gamma$-detectable pulsars.
Indeed, one misses producing the observed $\gamma$-ray pulsar numbers by a very
large factor unless the birthrate is nearly 4 times that allowed by the
observed OB stars and supernovae (i.e. more than $10\sigma$ higher than the value
estimated by \citealt{li10}). Accordingly, the birth spin period distribution,
for the $\gamma$-ray pulsars at least, must extend to much lower values.

	It is, however, possible to reconcile this with the radio results.
We find that the requirement for a large $P_0$ in the radio sample
is a consequence of the assumed radio luminsosity law.
We consider here three radio pseudo-luminosity ($L=S\,d^2$) laws. The first is
a power law distribution that is independent of any other pulsar
parameters \citep{lor06}.  Next, \citet{fg06} recommend a log-normal 
distribution whose central value scales as a power law with spin-down luminosity.
Finally, we consider the broken power law distribution of \citet{stoll87}
for which the power law scaling of the central value stops and is flat
above a cutoff (at $\dot{E} \approx 10^{34}\,{\rm erg\ s}^{-1}$).
We will refer to these functions as $\rm L_R$ Flat,
$\rm L_R$ Power Law, and $\rm L_R$ Broken Power Law, respectively.

\begin{figure}[h!]
\vskip -0.5truecm
\hskip -0.5truecm
\includegraphics[scale=0.47]{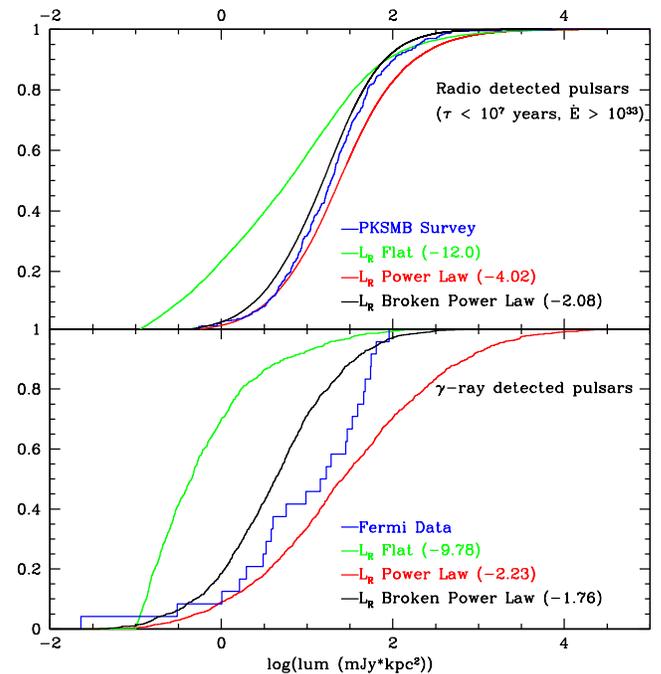}
\vskip -0.5truecm
\caption{\label{RadioLuminosity} Upper panel: Cumulative distribution functions
of radio luminosities for a population of young radio-detected pulsars. 
The model CDFs show radio luminosity laws with differing spin-down
dependence of the central value: Flat (green), Power Law (red) and 
Broken Power Law (black).  These models are compared with the Parkes Multibeam sample (in blue);
lower panel: same as above, but for the $\gamma$-ray-detected set,
with the LAT sample in blue.  The labels show the ${\rm Log_{10}}$
of the KS probability of agreement for each case.}
\end{figure}

	To illustrate the effect of the initial spin and radio luminosity
laws, we compare with the young pulsar populations: the LAT pulsar sample 
(all sky) and the high energy ($\dot{E}\ >\ 10^{33}\,{\rm erg\ s^{-1}}$) 
PMBS detections (within the PMBS footprint). To guage pulsar detectability
we adopt a standard radio beam width 
\begin{equation}
\rho=5.8\deg P^{-1/2}
\end{equation}
\citep{r93}, with the assumption of a circular beam of uniform
(integrated) radio flux directed along the magnetic dipole axis;
we discuss briefly later the possibility that young
objects have even wider radio beams from high altitude \citep{kj07}.
We simulate the Galactic pulsar population as summarized above,
assign a pseudo luminosity, check pulsar detectability at the modeled
beam orientation and distance, compute the cumulative radio luminosity 
functions of the model
detections (Figure \ref{RadioLuminosity}), and compare with the observed
samples. In all cases the `Flat' luminosity law does not reproduced 
the observed luminosities, so we discard this form. The broken power 
law and power law distributions are
quite similar for the radio sample (upper panel), but differ for the more energetic
radio-detected pulsars in the $\gamma$-ray sample. We remind the reader
that we have not followed the detailed selection effects of the radio
surveys. However, these are unlikely to affect these conclusions.
As an example, we applied a roll-off of the sensitivity
with period, as in \citet{det85}.
Here $S_{min} = S_0 [9w_e/(P-w_e)]^{1/2}$ with the effective
pulse width $w_e = [(0.1P)^2 + 10{\rm \, ms}]^{1/2}$ to approximate the 
PMBS losses at $DM\approx 200\, {\rm pc\, cm^{-3}}$.
This caused only a small
($\sim 6\%$) decrease in the number of detections for the most energetic 
pulsars (${\rm Log}({\dot E}) > 36.5$) and negligible loss 
($<$2\%) from the sample as a whole. In particular, we recover the \citet{fg06}
result that, with the Power Law model, $P_0=300\,$ms provides a very good
match to the PMBS radio luminosity function.

\begin{figure}[b!!]
\vskip -0.2truecm
\hskip -0.2truecm
\includegraphics[scale=0.45]{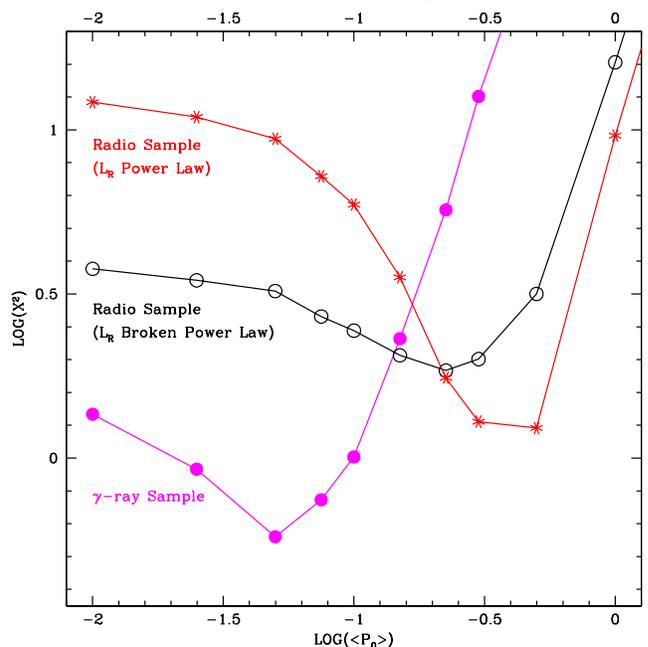}
\vskip -0.5truecm
\caption{\label{InitialPeriods} Goodness-of-fit of the model to
the observed data as a function of initial spin period.  For each model,
the pulsar population is assigned an initial spin period chosen from a
normal distribution with centroid and $\sigma$ $P_0$.
Magenta curve (filled circles) -- fit to the $\gamma$-ray pulsar 
sample. Red curve (starred points) -- the young radio pulsars, using
a Power Law radio luminosity law. Black curve (open circles) --
radio pulsar comparison with the Broken Power Law luminosity function.
}
\medskip
\end{figure}

	However, we see that the {\it Fermi} sample (Figure \ref{RadioLuminosity},
lower panel) shows some discrimination between the models, especially
for the high luminosity pulsars which dominate this set. The numbers of
high ${\dot E}$ pulsars are particularly sensitive to $P_0$. Accordingly 
we have explored this sensitivity by comparing the modeled
$N(\dot{E})$ of pulsars in the energetic ($\dot E > 10^{33}\,\rm erg\, s^{-1}$)
PMBS and {\it Fermi} samples as we vary $P_0$.
Our statistic is a $\chi^2$ comparison of the pulsar detections
in half-decade bins of $\dot{E}$. We also use the additional measurement
of the Galactic core-collapse rate (and error estimate) from \citet{li10},
fitting to the total birthrate. The results are shown in Figure \ref{InitialPeriods}.

	As expected, this shows that with the Power Law luminosity model the 
radio sample prefers long initial periods, but $P_0 = 300$\,ms is completely 
unacceptable for the $\gamma$-ray sample. This is because of the very small
birthrate of energetic pulsars for this large $P_0$. One can, of course, 
imagine a $\gamma$-ray emissivisity law allowing high efficiency for detection
of these very rare pulsars. The cost is that such pulsars are then extremely 
luminous and seen at very large distance. We find that the $\gamma$-ray pulsar
sample gives such a sensitive probe of the high ${\dot E}$ population that
changes sufficient to accomodate $P_0$ as large as 300\,ms predict a pulsar
sample in strong disagreement with the observed population
(see further discussion in \S 4 for a specific example).

   However, if we adopt the Broken Power Law radio luminosity law, we find the
prefered value of $P_0$ for the PMBS sample decreases and that the
$\chi^2$ at the $P_0 \approx 50$\,ms demanded by the $\gamma$-ray 
pulsars shows only a small increase from the minimum value. We suspect
that a complete treatment of the radio survey selection effects which should
decrease the detectability of short $P$ radio pulsars along the Galactic
plane would further improve the agreement at small $P_0$.

	In sum the radio and $\gamma$-ray pulsars can come from the same
population if $P_0 \approx 50$\,ms and the extreme increase in radio
luminosity for the shortest period pulsars is mitigated with the
Broken Power Law. A second quantitative comparison of this agreement is shown 
in Figure \ref{RadioLuminosity}, where we quote the model-data
Kolmogorov-Smirnov comparison for the two pulsar samples for each
luminosity law (here $P_0 = 50$\,ms). The Broken Power Law indeed provides
the best match to the detected radio luminosity function, especially
for the PMBS sample. For the smaller {\it Fermi} sample the
improvement is not large. We have confirmed that both this
test and the $N(\dot{E})$ test are insensitive to the chosen 
$\gamma$-ray model (see Section \ref{PulseMorph}).
We thus adopt the `Broken' luminosity law as the best fit to the data.

	As a consistency check, applying our adopted beaming and flux 
law to the population, we find that the
observed number of PMBS pulsars corresponds to a full Galactic
pulsar birthrate of $1.43/ {\rm 100\,yr^{-1}}$. This is in
reasonable agreement with earlier radio population estimates and with
the core-collapse estimates in \S \ref{GalStruct}
so we may turn our attention to the $\gamma$-ray-detection criterea.

\subsection{$\gamma$-ray Emission }

  To complete the population synthesis we require a $\gamma$-ray pulsar 
flux and beaming model. Such models also depend on the pulsar's energetics
and geometry and a major goal of the data comparison is to constrain
such dependence.  All of the $\gamma$-ray models employed in this work 
utilize the same heuristic efficiency law
\begin{equation}
\label{Efficiency}
\eta=(10^{33} {\rm erg\, s^{-1}}/{\dot{E}})^{1/2},
\end{equation}
with $\gamma$-ray production ceasing for objects with
$\dot E < 10^{33}\,\rm erg\,s^{-1}$. This law has theoretical
motivation \citep{a06} and provides a reasonable match to the observed
pulsar efficiencies \citep{psrcat}.
The portion of the magnetosphere producing $\gamma$-rays
and hence the fraction of the sky covered by the beam, however, differs
appreciably between models.

	In this paper we treat several versions of the two popular 
vacuum magnetosphere beaming models.  The first is the standard 
Outer Gap (OG) model, with emission in each hemisphere produced by
field lines above a single pole. In its basic form the $\gamma$-ray
production is dominated by the upper boundary of a vacuum gap
which is bounded by the last closed field lines and runs from the
null charge surface ($\Omega \cdot B = 0$) to the light cylinder 
$R_{LC}=cP/2\pi$. This gap has a spindown-dependent thickness
approximated by
\begin{equation}
\label{OGw}
w=\eta=({10^{33}{\rm erg\ s}^{-1}}/{\dot{E}})^{1/2}
\end{equation}
where w is the fractional distance from the last closed field
lines to the magnetic axis,
and the emission peaks $\sim w$ away from the closed zone.
As the pulsar spins down, $w$ grows; this causes the radiating particles to approach
the central field lines, the null charge crossing to move to high altitudes,
and the $\gamma$-ray emission
to become more tightly confined to the spin equator. To compute light curves
from this simple version, we compute photon emission from particles
traveling in a sheet centered on $w$, with a Gaussian
cross section. As ${\dot E}$ decreases, we might expect pair production
to become more difficult and this sheet to thicken. An extreme
assumption is that emission arises from particles spanning the
full thickness $w$ inward from the closed zone. Models computed
for this assumption are denoted full gaps ($\rm{OG_{FG}}$). In both
versions, this is essentially a `hollow' cone model which tends
to produce two caustic peaks, of varying separation, with a 
bridge between.

	The other popular beaming model has emission extending from
the star surface to high altitude, so that both poles are visible 
in both hemispheres. By truncating the emission somewhat before
the light cylinder it is possible to produce models lacking the
leading `OG' pulse. This picture most easily produces pulses
with $\sim 180^\circ$ phase separations and tends to have `Off pulse'
flux comparable to the `Bridge' between the two peaks. In its
original form this Two-Pole Caustic (TPC) model \citep{dr03}
truncates emission at the lesser of $r_{\rm{LC}}$ from the star or 
a distance of $0.75r_{\rm{LC}}$ from the spin axis.
This basic model uses the same $w$ relation as the OG model
(Equation \ref{OGw}).

	Modifications of this geometry have been suggested
as better approximations of the physical realization of such
vacuum zones in the Slot Gap model \citep{mh04,get10}.
For the first version (TPC2), the emission extends to 
to $0.95r_{\rm{LC}}$ and utilizes a slower, plausibly more 
physical, $w$ scaling:
\begin{equation}
\label{TPC2w}
w=\frac{1}{2}\,({10^{33}{\rm erg\ s}^{-1}}/{\dot{E}})^{1/6}
\end{equation}
which has wide $w > 0.1$ gaps even for very energetic Crab-like
pulsars.  Finally, we also define a full gap version of
this model ($\rm{TPC2_{FG}}$), with $\gamma$-ray emission coming
from the full volume between the last closed field lines and the
$w$ value calculated from Equation \ref{TPC2w}.

	There are also now a promising set of non-vacuum models,
generated by numerical realizations of the force-free magnetosphere.
One version, the `Separatrix Layer' (SL) model \citep{bssl} generates
emission from field lines that start from $w\approx 0.1-0.15$ 
and merge in the wind zone. In this picture, $\gamma$-ray emission 
arises from near the light cylinder, extending several 10's of
percent past this distance. While this model has some attractive
features, especially for young pulsars with robust pair production, 
it requires numerical realizations. We concentrate here on comparing
the {\it Fermi} sample with the analytically-derived vacuum models
(OG and TPC) and defer comparison with the SL model to future work.

\subsection{Pulsar Samples and Detection Sensitivity}
\label{Samples}

	The most uniform $\gamma$-ray pulsar sample available at present
is the set of 38 young objects reported in the First {\it Fermi} LAT Catalog of 
$\gamma$-Ray Pulsars \citep{psrcat}. We will refer to this as the ``6 Month'' 
sample (because that work was based on 6 months of {\it Fermi} LAT data).  
We also consider a larger sample of 50 young $\gamma$-ray pulsars,
including all non-recycled pulsar detections announced as of July 2010. 
In particular, this includes the `blind search' detections described 
in \citet{sp10}, as well as several individually announced discoveries. 
We refer to this as the ``Current'' sample; typically $\sim 1$ year of 
LAT exposure contributed toward these detections.

	{\it Fermi} has two different routes to $\gamma$-ray pulsar
discovery. When the signal is folded on an ephemeris, based typically
on radio observation, significant pulsations can be seen to
quite low levels. We term these ``radio-selected'' pulsars. Pulsations
can also be discovered `blindly', by searching directly in the GeV photons,
but this method requires a somewhat higher flux for detection. We use here
a threshold increase of $3 \times$ for such ``$\gamma$-selected''
detections \citep{psrcat}.

	A simple division of pulsars into these two classes is 
complicated by the discovery of several $\gamma$-ray pulsars with radio
luminosities an order of magnitude or more fainter than those of the typical 
pulsar population \citep{c09, a10}.
These sources were detected in deep, targeted integrations of young supernova 
remnants, unidentified $\gamma$-ray sources or detected $\gamma$-ray pulsars.
Their low flux density would not be accessible in the
typical sensitivity of modern large area sky surveys, 0.1 mJy at 1.4 GHz.

	Since we are simulating the properties of sources detectable in large
sky surveys, we wish to assign real detected pulsars to one of these classes
based on their observed fluxes, not the accident of whether they had
exceptionally deep radio observation. Thus ``radio-selected'' objects must have 
the survey-accessible flux density above. $\gamma$-selected objects 
must have the larger GeV flux required for `Blind search' discovery. 
This results in a re-classification of several objects.
There are three pulsars that had prior radio detections with 1.4\,GHz flux 
densities below 0.1 mJy (PSR J0205+6449, 0.04 mJy; PSR J1124-5916, 0.08 mJy;
and PSR J1833-1034, 0.07 mJy).  For the purposes of this work, we do not consider these
objects to be radio-selected $\gamma$-ray pulsars.  Two of the three objects have
high enough $\gamma$-ray flux levels that even without the assistance of a
radio ephemeris-guided folding search, they would still have been detected as
$\gamma$-selected pulsars (PSR J0205+6449 and PSR J1833-1034).
Thus, these two pulsars are added to the $\gamma$-selected subset of the observed population.
The third object, PSR J1124-5916, likely could not have been detected to date in 
blind $\gamma$-ray searches, and so is removed from the observed population.
Finally, PSR J2032+4127 was found in a blind $\gamma$-ray search, but radio
follow-up observations found a 1.4 GHz flux of 0.24 mJy.
Thus this object ``should have'' been detected in radio surveys
and is added to the radio-selected subset of the observed population.

	With these classification amendments, the ``6 Month'' sample  of
37 young $\gamma$-ray pulsars contains 19 radio-selected and 18 $\gamma$-selected
pulsars. Similarly the ``Current'' sample contains 24 radio-selected and
26 $\gamma$-selected objects.
These are the samples against which we will test the $\gamma$-ray emission models.

	In our simulation, any $\gamma$-ray pulsar with a radio beam crossing the Earth
line-of-sight with modeled flux density $> 0.1\,$mJy is assumed detectable in
a survey and termed radio-selected. For the $\gamma$-ray pulsars we
model the phase averaged flux on the Earth line-of-sight and compare with the
estimated detection threshold for each sky location plotted as a map in
\citet{psrcat}. We scale the ephemeris detection threshold with the square root
of exposure time from the 6-month estimates in that paper. To be $\gamma$-selected,
a modeled pulsar must exceed $3\times$ this ephemeris detection threshold,
{\it and} must have a modeled radio flux density $< 0.1\,$mJy at Earth.

\section{$\gamma$-ray Pulse Morphology}
\label{PulseMorph}

	In addition to simple detections and luminosities, the
observed radio and $\gamma$-ray pulsar profiles contain a wealth
of information on the pulsar emitter. We have shown in \citet{rw10}
how detailed analysis of the light curves, radio polarization, and other
multiwavelength data can make strong model comparison statements for
individual sources.  Here we wish to treat the statistics of the population
as a whole, so we must rely on relatively crude basic pulse 
properties that can be measured relatively uniformly in the
discovery data. We thus characterize the pulses by $i$) the number
of strong, caustic-type peaks in the $\gamma$-ray light curve
$ii$) the phase differences between the radio and $\gamma$-ray peaks
and $iii)$ the relative strength of $\gamma$-ray emission between
(`bridge') and outside of (`off') the peaks for the common double-peaked
profiles. In the following sections we quantitatively compare each
of these {\it Fermi} observables with the model predictions for the pulsar
population.

\subsection{Peak Multiplicity}

	The first characteristic typically measured for a light curve is
the number of peaks in the pulse profile. To compare with the data,
we have taken each pulsar in our simulated Galaxy, applied the beam shape
for each of the five trial $\gamma$-ray models described above, and tagged peaks
with an algorithm that identifies peaks and their phases in the
modeled $\gamma$-ray light curves. To increase similarity to the
actual data we blocked the model light curves into 50 phase bins,
before running the peak tagging algorithms.
The results range from 0 peaks ($\gamma$-ray flux on the Earth line-of-sight,
but no strong peak detected) to 4 or more distinct sharp peaks.
Since for some models the peak morphology varies significantly
with pulsar age/spin-down luminosity, we divide the sample into
4 bins of ${\dot E}$, logarithmically spaced.

\begin{figure*}[t!!]
\vskip -0.7truecm
\hskip -0.4truecm
\includegraphics[scale=0.98]{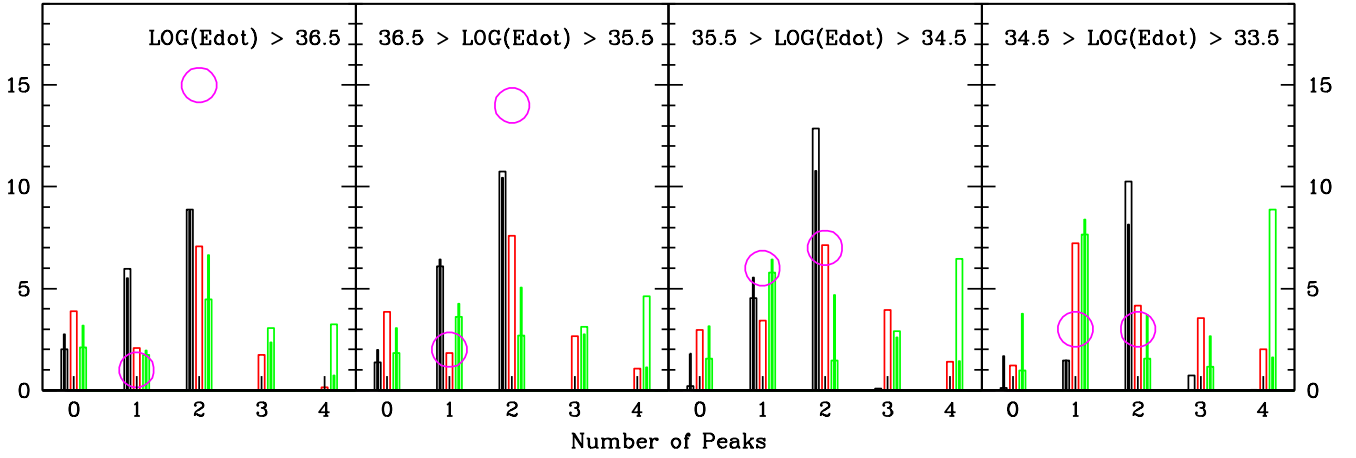}
\vskip -0.3truecm
\caption{\label{NumPeaks} Peak multiplicity histograms in four different
$\dot{E}$ bins.  The black bars are for the OG model, the red is for the
Dyks and Rudak TPC model, and the green is for the TPC2 model. The
distributtion of observed pulsars in each bin is given by the magenta
circles.
The thick hollow bars use a narrow Gaussian distribution of field lines
around the w value specified from Equation \ref{OGw} or \ref{TPC2w}, appropriately.
The thin filled bars use the
full gap, stretching from the last closed field lines in to the field
lines specified by the appropriate w value.}
\end{figure*}

The results are shown as histograms in Figure \ref{NumPeaks}, with bars
for each of the five models. For comparison, the measured multiplicities
(always 1 or 2) for the ``Current'' sample are shown by the circles in each panel.
It is immediately apparent that the original TPC model (central bars) 
produces many 3 and 4 peak light curves not seen in the data. The amendments
to a slower $w$ evolution (Equation \ref{TPC2w}) and gaps extending 
to 0.95$R_{LC}$ (wide right bars) make this disagreement worse. Extending 
the emission throughout the full gap (narrow right bars) does blur out some weaker peaks, 
mitigating the problem. We will use this blurred version
(slow $w$ evolution, fully illuminated gaps) in the remainder
of the paper when we refer to `TPC', as it and the original TPC model
give the best match to the observed quantities. The differences in the outer gap
OG predictions between the Gaussian illumination (wide left bars)
and the full gap illumination (narrow left bars) is relatively
minor.

	We quantify the data comparison in Table \ref{PeakTable}
and Table \ref{MorphTable}. Rows 1 through 4 of Table \ref{PeakTable} 
show the Poisson probability that the model is a fit to the {\it Fermi} 
data.  Each row represents one energy bin, and the best fit model for 
that bin has its probability in bold. This emphasizes that the TPC2 
model fairs particularly poorly without the full gap illuminated,
and that for each bin the OG models provide substantially better 
representations of the peak multiplicity. The modeled presence of 
0 peak sources (which could not, by definition, be discerned in the data)
lowers the probability for all models. Also the absolute normalization is
meaningful here, as these computations are run with the same 
model-determined pulsar birthrate (see \S \ref{Conclusion}).
In row 5 we calculate the probability decrement of each model when compared
to the best model (original OG),
summed across the four energy bins.
Again, in the rest of the paper
we adopt the original OG and $\rm{TPC2_{FG}}$ version
as the representatives of the two fiducial models.

\begin{deluxetable}{rrrrrr}
\tablecolumns{6}
\tablewidth{0pc}
\tablecaption{Model Probabilities from Peak Number Comparison\label{PeakTable}}
\tablehead{
  \colhead{} & \multicolumn{5}{c}{log(Poisson Probabilities)} \\
  \cline{2-6} \\
  \colhead{$\dot{E}$} & \colhead{OG} & \colhead{$\rm OG_{\rm FG}$} & \colhead{TPC}
    & \colhead{TPC2} & \colhead{$\rm TPC2_{\rm FG}$} }
\startdata
$\rm{LOG}(\dot{E})\ > \ 36.5$            & {\bf -4.46} & -4.66 & -5.58 & -8.42   & -6.02       \\
$36.5\ > \ \rm{LOG}(\dot{E}) \ > \ 35.5$ & {\bf -3.17} & -3.59 & -5.82 & -11.22  & -7.38       \\
$35.5\ > \ \rm{LOG}(\dot{E}) \ > \ 34.5$ & {\bf -2.54} & -2.75 & -5.52 & -8.79   & -4.95       \\
$34.5\ > \ \rm{LOG}(\dot{E}) \ > \ 33.5$ & -3.43 & {\bf -3.19} & -4.97 & -7.03   & -5.80       \\
$\Sigma \Delta {\rm Log}(P)$ &  \nodata & -0.59 & -8.29 & -21.86 & -10.55 \\
\enddata
\end{deluxetable}

\begin{deluxetable*}{rrrrrrrrr}
\tablecolumns{9}
\tablewidth{0pc}
\tablecaption{Model Probabilities from Pulse Morphology Comparisons\label{MorphTable}}
\tablehead{
  \colhead{} & \multicolumn{2}{c}{$\Delta$ - $\delta$ : 2-D KS} & \colhead{} &
    \multicolumn{2}{c}{$\Delta$ : 1-D KS}  & \colhead{} &
    \multicolumn{2}{c}{Bridge/Off-pulse : FoM} \\
  \cline{2-3} \cline{5-6} \cline{8-9} \\
  \colhead{$\dot{E}$} & \colhead{OG} & \colhead{$\rm TPC2_{\rm FG}$} &
    \colhead{} & \colhead{OG} & \colhead{$\rm TPC2_{\rm FG}$} &
    \colhead{} & \colhead{OG} & \colhead{$\rm TPC2_{\rm FG}$} }
\startdata
$\rm{LOG}(\dot{E})\ > \ 36.5$            & {\bf -0.96}  &   -3.40  & &
  {\bf -2.00} & -2.05        & & {\bf 600 $\pm$ 128}  & 462 $\pm$ 142 \\
$36.5\ > \ \rm{LOG}(\dot{E}) \ > \ 35.5$ & {\bf -0.92}  &   -3.52  & &
  -0.97       & {\bf -0.72}  & & {\bf 1928 $\pm$ 268} & 387 $\pm$ 143 \\
$35.5\ > \ \rm{LOG}(\dot{E}) \ > \ 34.5$ & {\bf -2.30}  &  -4.52   & &
  -1.18       & {\bf -0.83}  & & {\bf 552 $\pm$ 198}  & 316 $\pm$ 159 \\
$34.5\ > \ \rm{LOG}(\dot{E}) \ > \ 33.5$ & \nodata      & \nodata  & &
  {\bf -0.65} & -1.52        & & {\bf 221 $\pm$ 138}  & 143 $\pm$ 116 \\
$\Sigma \Delta {\rm Log}(P)$ & \nodata & -7.26 && \nodata & -0.32 && \nodata & -7.37 \\
\enddata
\end{deluxetable*}

\subsection{Peak Locations}

\begin{figure*}[t!!]
\vskip -0.3truecm
\hskip -0.4truecm
\includegraphics[scale=0.98]{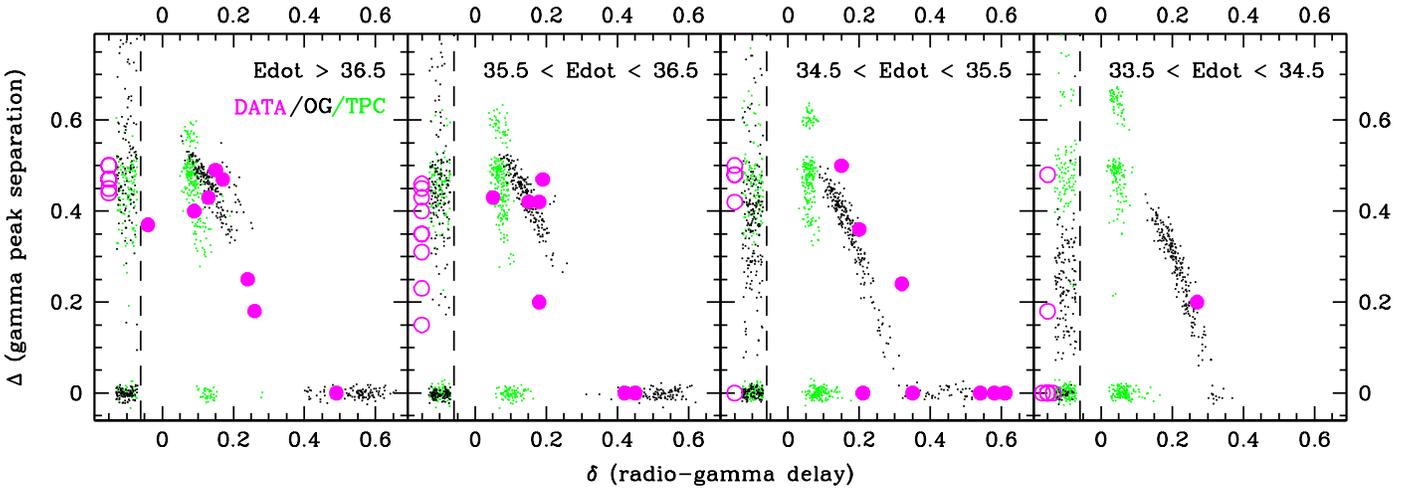}
\vskip -0.3truecm
\caption{\label{Deltadelta} 2-D plots in the $\Delta-\delta$ phase space,
for the same four $\dot{E}$ bins as in Figure \ref{NumPeaks}.}
\end{figure*}

	A particularly powerful test of the models arises from
comparing the $\gamma$-ray peak separations $\Delta$ (available
for all light curves with two or more peaks, taken as the widest strong peak
separation) and the lag of the first $\gamma$-ray peak from the
magnetic dipole axis $\delta$. For radio-detected pulsars this 
axis is typically marked by the main radio pulse. The
``$\Delta-\delta$'' plot of these quantities is an excellent
probe of the model geometry \citep{ry95,wat09}.
We produce such plots here in Figure \ref{Deltadelta}, again with 
four different $\dot{E}$ bins.

	$\gamma$-ray pulsars that are not detected in the radio will have
an undefined value of $\delta$; accordingly these objects are plotted in
a band to the left, beyond the vertical dashed line at
$\delta\sim-0.06$.  Similarly, pulsars with only a single peak in their
$\gamma$-ray pulse profiles have an undefined value of $\Delta$.
We assign to these singly-peaked profiles a value $\Delta=0$ and 
plot in a band along the x axis. Note that in practice limited
signal-to-noise in the $\gamma$-ray  profiles usually prevents
measurements of peak separations any closer than $\Delta \approx 0.2$,
so the observed (large dots) single pulse profiles doubtless
include some unresolved tight doubles, such as appear in the
models particularly for small ${\dot E}$.

	In the plots the data show a clear inverse correlation 
between $\delta$ and $\Delta$. This was predicted for outer
magnetosphere models \citep{ry95} and indeed appears for the (dark)
OG model points.  The trend is weak or absent in the (light) TPC points.
In the OG case the correlation is easily understood since both
caustic peaks are generated from the magnetic pole opposite to that
producing the radio emission. As the Earth line-of-sight cuts 
a smaller chord across this hollow cone, $\delta$ increases and 
$\Delta$ decreases. In contrast, the TPC model produces the
first $\gamma$-ray peak from the same pole as the radio emission.
The $\delta$ is nearly fixed, showing only small variation
with viewing angle and age.

	A less obvious difference is the OG correlation of
$\Delta$ and $\dot{E}$, again weak or absent in the TPC model,
which as expected shows a typical $\Delta =0.4-0.6$, with a
strong concentration at  $\Delta\approx 0.5$, for all spindown
luminosities.  In the LAT data, if we consider for each energy bin
the fraction of objects
with $\Delta>0.4$ we find 73\% (11/15), 56\% (9/16), 38\% (5/13), 
and 17\% (1/6), moving through our four energy bins from high to low.

	For a quantitative measure of the goodness of fit between
the data and the models, we employ a two dimensional Kolmorgorov
Smirnov test (KS test) in this plane on the radio-selected samples.
We can also run a standard one dimensional KS test on the $\gamma$-selected
objects, for which only a $\Delta$ value exists.  The probabilities 
from these tests are displayed in Table \ref{MorphTable}.
With only a single detection, the 2-D KS test is unavailable for the
lowest $\dot{E}$ bin. The bottom row of Table \ref{MorphTable} (as for Table 
\ref{PeakTable}) gives the summed probability decrement compared
to the best (OG) model; although the OG model itself is not adequate
(Prob $\sim 10^{-4}$), probabilities for the TPC models are factors of
$\ge 10^7$ lower. Statistically, the OG model is strongly preferred.

	Single peaked (x-axis) $\gamma$-ray light curves deserve some
additional discussion. As noted above, objects may appear here when
the data do not resolve a double. Such objects in the OG picture
are expected to appear at $\delta\approx 0.3-0.4$. In the OG model,
a peak may also be single if the first $\gamma$-ray peak is missing.
Since this caustic forms at high altitudes, it is rather sensitive
to field line distortion from sweep back and to aberration effects,
especially for moderate to large $\dot{E}$.  In the vacuum models,
we find that this caustic (and peak) are often missing for
energetic pulsars when the observer line-of-sight lies well away
from the spin equator. Near the spin equator the caustics are 
unaffected, so that large $\Delta$ appear, but smaller
$\Delta < 0.3$ are missing for energetic pulsars. As a result
$\delta$ is large ($\sim 0.5-0.6$) since it now measures the distance to
what is normally the second pulse. This effect may be seen in the models and data of
the left panels of Figure \ref{Deltadelta}. However, it should
be noted that the sensitivity of the first peak caustic to
field line perturbations makes the extent of this effect difficult
to predict. For example \citet{rw10} found that open zone currents
can reduce or eliminate such first peak `blowout', so realistic models
containing plasma may affect the prevalence of large $\delta$.

	Finally, it should be noted that {\it if} small altitude
emission occurs, then the trailing side of the radio-producing
pole may form a caustic (this is the first peak in the TPC picture).
As seen by the heavy concentration of TPC model dots in
Figure \ref{Deltadelta}, this results in $\delta \approx 0.1$ .
The {\it Fermi} LAT may have in fact uncovered one such object:
PSR B0656+14, which appears near $\delta = 0.2$ in the third
panel, has a single pulse, a peculiar soft spectrum and an
apparent luminosity $\sim 30\times$ smaller than seen from 
similar $\dot{E}$ pulsars. Polarization modeling suggests that
it has a very small $\alpha$ and $\zeta$, so that its outer
magnetosphere beam should miss the Earth. In this interpretation
we only see the fainter low altitude emission because of
this pulsar's very low distance.

	Thus $\delta$ provides a useful way of sorting pulsar
properties, even when only a single $\gamma$-ray peak is available. Improved
S/N, polarization studies and, above all more physical magnetosphere
modeling, should help us calibrate this diagnostic.

\subsection{Bridge and Off-pulse Emission}

\begin{figure*}[t!!]
\vskip -0.3truecm
\hskip -0.4truecm
\includegraphics[scale=0.98]{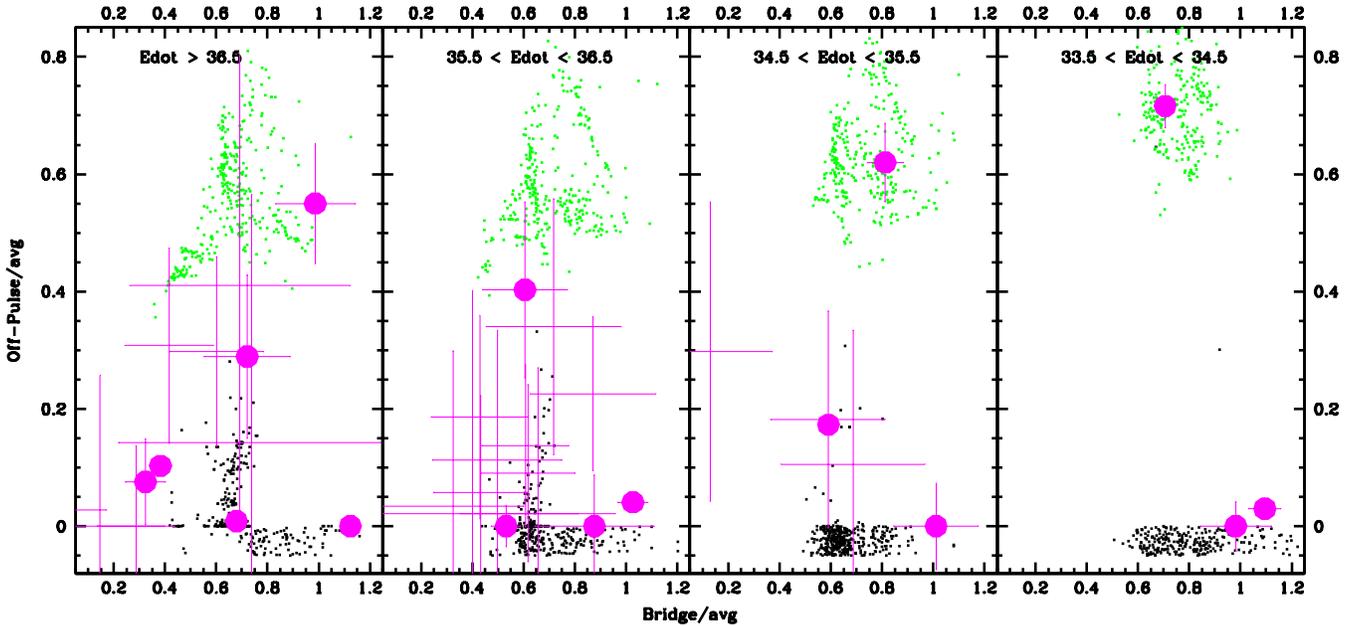}
\vskip -0.3truecm
\caption{\label{BridgeOffpulse} 2-D plots in the bridge emission $vs.$
off-pulse emission phase space, for the same four $\dot{E}$ bins 
as in Figure \ref{NumPeaks}. Estimated measurements for observed
double-peaked pulsars are plotted, assuming a 10\% uncertainty in
the background levels reported in the discovery papers. Only
objects with high ($>2\sigma$) significance have dots 
associated with the plotted error bars.
}
\end{figure*}

	The most common $\gamma$-ray profiles contain two peaks,
and for these profiles
we can introduce a third measure of pulse shape that can be applied to 
the bulk sample: the strength of the intra-peak (`bridge') and
inter-peak (`off-pulse') emission.  To estimate these flux
levels we divide the double peaked profiles into 4 phase windows. 
Two intervals of $\Delta \phi=0.14$ (7 bins in a 50 bin light curve)
are centered on the main peaks and the two remaining phase intervals
(totaling 0.72 of pulse phase) are assigned to the `bridge' and `off-pulse'.
We next measure the {\it mean} flux in each of these phase windows.
We then report the bridge fraction as 
$\langle$bridge$\rangle$/$\langle$(bridge+peaks)$\rangle$ and
the off-pulse fraction as 
$\langle$off-pulse$\rangle$/$\langle$(bridge+peaks)$\rangle$. 
This easily defined measure generally corresponds well to
a visual definition of bridge and off pulse intervals, although it
does not always isolate well the absolute pulse minimum.
We plot these two fluxes for the model light curves, in the
usual four panels of spin-down luminosity, in Figure \ref{BridgeOffpulse}.

	Clearly these quantities provide good model discrimination.
Models radiating only in the outer magnetosphere (e.g. OG) have little
off-pulse flux, especially for older pulsars, while models radiating
at all altitudes have comparable bridge and off-pulse emission. 
In particular, the TPC picture produces essentially no light
curves with off-pulse fluxes less than 40\% of the pulse phase emission.

	Unfortunately, unlike the peak number and separation, such
fluxes are not provided in \citet{psrcat}, \citet{sp10} and the 
other discovery papers. One difficulty in making such measurements
is the identification of the true background level. In practice
it can be very difficult to distinguish unpulsed ``DC''
magnetospheric emission from the contribution of an unresolved
background in close proximity to the pulsar, such as expected
from a surrounding pulsar wind nebula. Detailed study of source
extension and spectrum at pulse minimum can help distinguish these cases,
but this goes beyond the pulsar catalog data. 

	Nevertheless, the published light curves do plot an 
estimated background flux level.  Accordingly we entered these 
light curves, defined the peak and intervening intervals exactly 
as above and measure the bridge and off-pulse flux levels. We
assume a systematic 10\% uncertainty in the reported background level,
and then propagate this and the Poisson uncertainties in the
bridge-, off-, and average-pulse fluxes to produce the two
flux ratios and errors defined above. In a few cases the 
background level defined in \citet{psrcat} were above the measured
pulse minimum. In those cases, we reassigned the background flux
to this level (and by definition the off-pulse flux was then 0).
In a number of cases, the very large background pedestal prevented
accurate measurements. While we show error bars for all estimates of double
pulse pulsars in Figure \ref{BridgeOffpulse}, we mark with large dots
only those measurements which had either a 2$\sigma$ significance
measure of both flux ratios, or a ratio less than 0.1 at
2$\sigma$ significance.

	Bridge emission varies from $\sim 0.3\times$ to 
$> 1\times $ the average pulse flux at all spindown luminosities,
although a few high ${\dot E}$ objects seem to have virtually
no emission away from the peaks. In contrast, the majority
of the objects have $0.3\times$ or less of the pulse emission in
the off-peak window. Most of the best measured pulsars show
very small values. Values of $\sim 0.1-0.3\times$ appear for
a number of the most energetic pulsars, but these measurements
are of poor statistical significance. In any case such objects
are most likely to show unpulsed contamination from associated PWNe,
supernova remnants and other diffuse, but unresolved emission.
Thus overall, the general distribution of measured values
matches better to the OG predictions, as shown in Table \ref{MorphTable}.

	Nevertheless, several objects are distinctly inconsistent with
emission from only high altitudes. In particular PSR J2021+4026
(panel 3) and PSR J1836+5925 (panel 4), show very strong off pulse emission,
which is spectrally consistent with magnetospheric emission
(hard with a few GeV exponential cut-off). These unusual objects
lie squarely within the TPC model zone. Interestingly both are
quite low ${\dot E}$, and both have $\Delta$ very close to 0.5,
consistent with a low altitude, two-pole interpretation.
Also, the very energetic PSR J1420-6048 (panel 1) shows strong 
off pulse emission. Here, with $\Delta =0.26$ a low altitude
interpretation is not natural. However, unlike the other two 
cases, the very large and poorly determined background (from
the bright PWN emission in the surrounding `Kookaburra' complex)
may exceed the catalog background flux level. However it is clear
that, at least in a few cases, an emission component other than the
outer magnetosphere of the OG model contributes substantially
to the pulse profile. An excellent candidate is low altitude
emission, such as posited in the TPC model. An alternative is a
current-induced shift of the outer gap start below $r_{NC}$
\citep{h06}. It will be particularly
interesting to discover the physical effect that causes such
emission to be significant for a few pulsars, but negligible for 
most.

\section{$\gamma$-ray Evolution with $\dot{E}$}
\label{Evolution}

	As illustrated in the four-panel Figures 
\ref{NumPeaks}-\ref{BridgeOffpulse}, the pulse profile properties 
evolve as the pulsar ages and ${\dot E}$ decreases. In addition
there is of course strong evolution in the pulsar luminosity.
Together, these trends strongly affect the pulsar detectability in the
two detections classes ($\gamma$-selected and radio-selected)
as a function of spindown flux. We explore these
population effects in this section.

\begin{figure}[h!]
\vskip -0.6truecm
\hskip -0.5truecm
\includegraphics[scale=0.45]{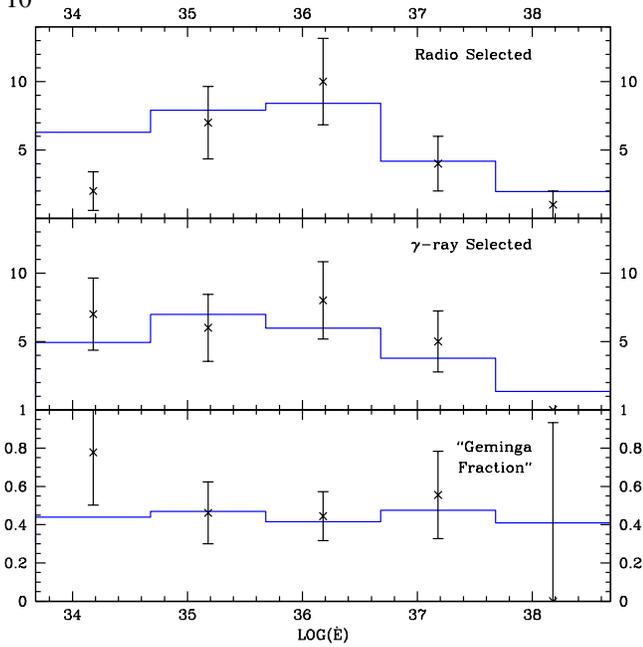}
\vskip -0.5truecm
\caption{\label{EdotHist} Histograms of detection numbers versus $\dot{E}$.
The model overproduces radio-selected $\gamma$-ray pulsars at low $\dot{E}$.} 
\end{figure}

In Figure \ref{EdotHist} we plot histograms of detection numbers versus
spin-down luminosity, for both the radio-selected (top panel) and
$\gamma$-selected (middle panel) subsets.
The blue histogram shows the simulated OG model detections, while the black
data points show the {\it Fermi} results, with statistical error bars.
In the bottom panel we plot the ``Geminga Fraction,'' or fraction of the total
number of detected pulsars which are $\gamma$-selected.

	The OG model gives a reasonably good fit to the {\it Fermi} data,
with $\chi^2$ Gehrels fits of 0.587 per degree of freedom for the
radio-selected population, 0.251 per degree of freedom for the
$\gamma$-selected population, and 0.054 per degree of freedom for
the Geminga Fraction.
\citet{ret10} have noted that in the 6 month sample
the most energetic $\gamma$-ray-detected pulsars are all radio-detected as well.
They argue that this implies 
a large radio beam, roughly co-located with the $\gamma$-ray emission region
for these most energetic (${\dot E} > 6 \times 10^{36} {\rm erg\, s^{-1}}$)
pulsars. This may be caused by high altitude emission as posited in \citet{kj07}.
Our model, without high altitude radio beams, predicts 7-8 such very high
$\dot{E}$ pulsars in the 6 month sample, of which 2-3 should be undetectable
in the radio (i.e. our line of sight is outside of the radio beam).
The actual LAT sample had 7 such pulsars, all of which were detected in the radio.
We have simulated high-altitude radio emission, confirming that it
prevents the $\gamma$-ray only detections at the highest $\dot{E}$. However,
we conclude that the present sample is too small to use these numbers
to probe the detailed behavior of such objects with high statistical significance;
for the moment the case for high altitude radio emission still largely relies on the
radio pulse properties. 

There does exist one moderately significant defect in the histogram for low
$\dot{E}$ radio-selected pulsars; the model predicts too many such detections.
This is reflected as well in the Geminga Fraction,
where the data demand a larger value for the Geminga Fraction at low energies.
Such an increase is not seen in the model, due to the excess of radio-selected
detections.  For now we note the discrepancy, but refer to Section \ref{Amendments}
for discussion of a possible solution.

\begin{figure}[h!]
\vskip -0.35truecm
\hskip -0.2truecm
\includegraphics[scale=0.45]{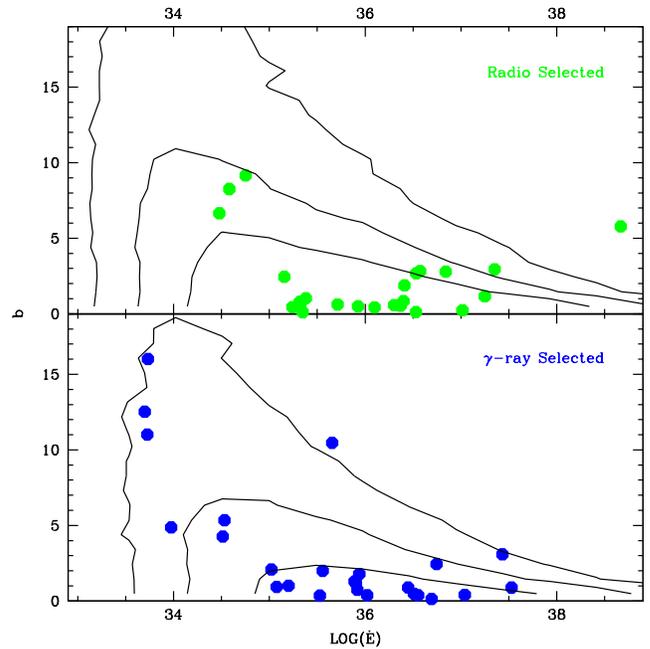}
\vskip -0.5truecm
\caption{\label{Edotvsb} 2-D plot of detections in the $\dot{E}$-galactic
latitude plane.  Note the overproduction of model detections at low $\dot{E}$
in the radio-selected sample.}
\end{figure}

	Spin-down luminosity should, through the $\gamma$-ray efficiency,
be correlated with the typical distance of the detected pulsars. Unfortunately,
as we have seen, distances estimates (and hence inferred luminosities) are
very poor for many of these objects. However, Galactic latitude correlates
inversely with distance and is a directly observable property. Accordingly,
in Figure \ref{Edotvsb} we show the ${\dot E} - b$ planes for the
radio-selected and $\gamma$-selected subsets.  The colored points show the
locations of {\it Fermi} pulsars and the contours show the density of the
OG model detections.

	The modeling predicts a dramatic increase in the latitude spread of
the lowest ${\dot E}$, (lowest $L_\gamma$, nearest) objects. The
$\gamma$-selected set actually shows this trend quite well. In contrast,
the radio-selected pulsars show a distinct lack of the nearby, lower luminosity
higher latitude objects predicted by the models at log$(\dot{E}) < 34.5$. 
This is directly connected to the simple lack of low $\dot{E}$ radio-selected
objects noted above. Also in the radio panel, one notices the very high 
$\dot{E}$ Crab pulsar -- which is surprisingly close for such an
energetic pulsar and notoriously far from the plane.
This distance off the plane is likely a product
of an unusually long-lived low mass runaway progenitor from the Gem OB1
or Aur OB1 associations.

	The model agreement is tested with a two dimensional KS test.
For the $\gamma$-selected sample one gets a KS probability of 3.9\%, 
quite acceptable for this population model. For the radio-selected sample
the KS probability is however only 0.5\%, which is not a good agreement.
The discrepancy is largely driven by the missing 
${\dot E} < 10^{35}\ \rm{erg\ s^{-1}}$ pulsars; if this region is excluded the
KS probability rises to 2.6\%, which is not excludable.

\begin{figure}[h!]
\vskip -0.35truecm
\hskip -0.2truecm
\includegraphics[scale=0.45]{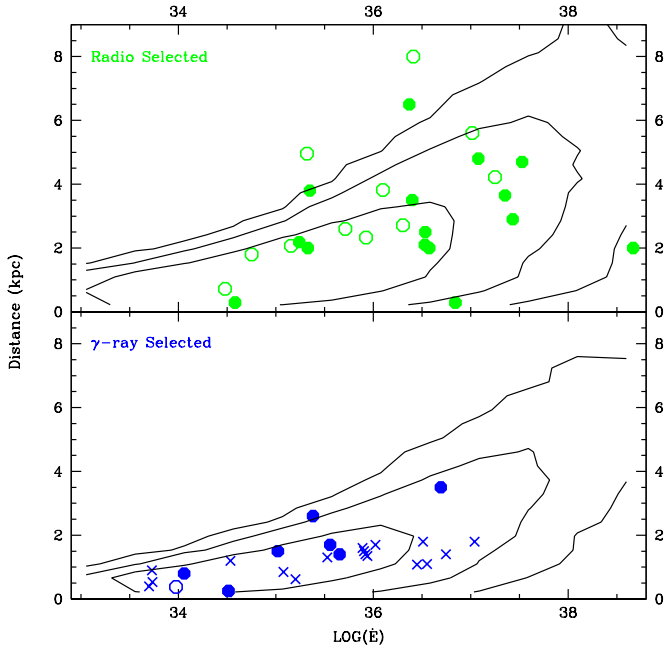}
\vskip -0.5truecm
\caption{\label{Edotvsd} 2-D plot of detections in the $\dot{E}$-distance plane.
The overproduction of model detections at low $\dot{E}$
in the radio-selected sample is still evident.}
\end{figure}

	With the caveat about observational uncertainties noted above
we can also compare with the data in the $\dot{E}$-$d$ plane (Figure
\ref{Edotvsd}). Here we expect small $d$ at low $\dot{E}$. In an attempt
to illustrate the distance uncertainties, we use different symbols
to plot objects with distance estimates from various sources (drawn
from the discussion in the discovery catalogs).  For example, radio
detected pulsars have Dispersion Measure (DM) distance estimates. These
are shown as open circles. While generally held to be accurate to 
$\sim 30$\% for the bulk of the pulsar population, these estimates
evidently have much larger fractional errors for the energetic
LAT pulsars near the plane and, in a number of cases, are clearly substantial
overestimates. In some cases we have other distance estimators
ranging in reliability from astrometric parallax (very high) through
HI absorption kinematics to spatial associations (low). These objects
are plotted with filled dots. Note that in the upper panel of
Figure \ref{Edotvsd} these distances average lower than the DM values for
all ranges of $\dot{E}$.

	Finally, for a number of $\gamma$-selected
pulsars we have no estimate of distance other than assuming
a $\gamma$-ray efficiency (Equation \ref{Efficiency}) and isotropic emission
and using the observed $\gamma$-ray flux to estimate $d$. Such
estimates are referred to as ``pseudo-distances'' \citep{sp10}, and while they
provide helpful consistency checks, their use is, to some degree,
circular if employed to test a model. They are plotted as crosses.

	Examining the 2-D KS test probabilities,
we find that the radio-selected sample returns a
probability of 0.7\%.  As for the $\dot{E}-b$
set, this is moderately excludable.  Again, a re-test using
only objects above
$10^{35}\ \rm{erg\ s^{-1}}$ delivers a higher probability of 2.3\%.
The probability for the $\gamma$-selected sample is 19.9\%. This large value
is certainly partly due to the use of the pseudo-distances.  However,
even after removing these objects, we find a probability of 10.4\%, 
suggesting that the model is a quite good representation of the data.

	Thus our heuristic $\gamma$-ray luminosity law (Equation 
\ref{Efficiency}) and our Broken Power Law radio-luminosity law provide a good 
match to the data. Other models provide a much poorer match.
For example, a simple linear $\gamma$-ray luminosity law $L_\gamma= 1/3 {\dot E}$\,
together with the straight Power Law radio luminosity law and large
birth period $\langle P_0 \rangle =300$\,ms recommended by \citet{fg06} can produce
a reasonable number of $\gamma$-detectable pulsars for a
\citet{li10} birthrate. However the predicted distances are $2\times\; -\; 4\times$
those observed and 2-D KS comparisons in the ${\dot E}-b$ and
${\dot E}-d$ planes produce unacceptably low probabilities for the
radio-selected $\gamma$-ray sample
($9 \times 10^{-4}$ and $7\times 10^{-4}$, respectively).

\section{Possible Amendments to the $\gamma$-ray Model}
\label{Amendments}

	We have shown how to quantitatively compare the observed
properties of the LAT pulsar survey to predictions of population 
models.  While the agreement is generally quite good, in particular
for the models dominated by outer magnetosphere radiation, some
discrepancies are already seen. Our ability to study such discrepancies
will certainly improve as the size of the LAT sample and the quality
of the individual light curve measurements increase. Thus we wish 
to discuss sensitivities of the observed data sample to details of 
the beaming model and amendments to the modeling that may further 
improve the fidelity of the synthesized population.
	
	As an example, we should consider how beaming and luminosity
evolve at low $\dot{E}$ as $w$ and the pulsar efficiency approach
a maximum and the gap saturates and shuts off. In particular,
for the OG model, since emission starts above the null charge
surface, large $w$ drives this start toward the light cylinder and
produces a decreasing $\gamma$-ray beam solid angle.

\begin{figure}[h!]
\vskip -0.5truecm
\hskip -0.2truecm
\includegraphics[scale=0.45]{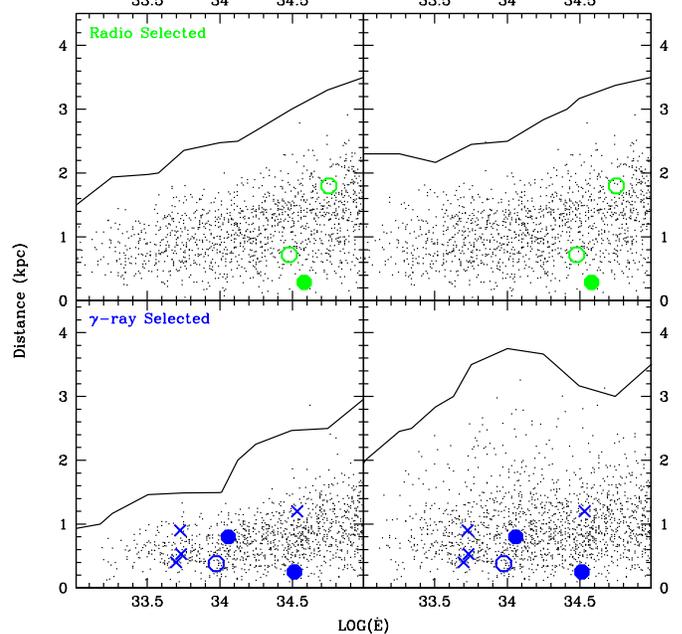}
\vskip -0.5truecm
\caption{\label{GapPower} Figure showing the effects of forcing the model
to pump $\eta\dot{E}$ power into the $\gamma$-ray emission region,
regardless of how small that region becomes.  Note that we have zoomed
in here on the region of interest, $\dot{E} < 10^{35} \rm{erg s^{-1}}$.}
\end{figure}

	If one follows the simple efficiency law (Equation \ref{Efficiency}) into
such a regime then phase average pulse intensity for the few observers 
seeing all of $\eta \dot{E}$ in this very small angle would become 
very large just before the beam shuts off. In principle, this allows
a small number of $\dot{E} < 10^{34}\,\rm erg\,s^{-1}$ pulsars to be seen
at very large distances. The more physical alternative, adopted
here, is to link $\eta \dot{E}$ to the `surface brightness',
the thickness-integrated emissivity per unit area of the gap volume. For
all ${\dot E}$ producing modest $w$ this means $L_\gamma \propto w$
or $\propto w^3$,
as usual. However as $w$ approaches unity and the active gap 
area contracts laterally, the total sky-integrated luminosity
smoothly goes to $0$ at gap saturation, although the surface 
brightness continues to grow. To illustrate the sensitivity 
of the data-population comparison to such effects we illustrate
(Figure \ref{GapPower}) the different prediction of these two cases.
The left panels reproduce the low ${\dot E}$ end of 
Figure \ref{Edotvsd}, while the right panels show the increased
distance reach  -- from $\lesssim1\,$kpc to $\gtrsim 3\,$kpc --
if all the power is forced into the decreasing beam. As it 
happens the effect is largely
concentrated in the $\gamma$-selected pulsars, since to produce
emission from such large $w$, $\alpha$ must be moderate to small
(i.e. $\alpha\lesssim50^\circ$), but the viewing angle $\zeta$ 
must be very near the equator (i.e. $\zeta\gtrsim85^\circ$).
For such objects the radio beam misses Earth. This
and even more subtle evolution effects will become increasingly
subject to test as the LAT sample grows.

	We have already alluded to another deficiency in our modeling
of the low  ${\dot E}$ population; the substantial over-prediction
of radio-selected objects. In fact, there is an evolutionary process 
that can be amended to our model to produce exactly this effect:
magnetic alignment.
Recent work has suggested that a secular decrease in $\alpha$ may 
occur on timescales as short as 1 Myr \citep{y10}.  Given the high
sensitivity of the $\gamma$-ray beaming to $\alpha$ for some models,
even partial alignment can have a large effect on 
our young pulsar population.

	We describe here qualitatively the main effects, deferring
detailed population predictions to a study including the alignment kinematics.
As $\alpha$ decreases with age, there are two main effects
on the pulse observables in the OG model: a decrease in the average $\Delta$
and a decrease in the number of total detections, with an especially strong
elimination of radio-selected objects. These should show up principally at
the largest ages (i.e. lowest $\dot{E}$ values) of the $\gamma$-ray pulsar population.

	The first OG effect, a decrease in typical $\Delta$ values, results
from  the fact that the highest $\Delta$ values are always produced
by large $\alpha$ models.  Eliminating such models would have the effect
of shifting the clump of $\Delta \sim 0.4$, $\delta \sim 0.15$ models
in Figure \ref{Deltadelta} panel 4 to  $\Delta \sim 0.2$, 
$\delta \sim 0.25$. Of course we need substantially more low ${\dot E}$
pulsar detections before any such evolution can be tested.

\begin{figure}[h!]
\vskip -0.5truecm
\hskip -0.2truecm
\includegraphics[scale=0.45]{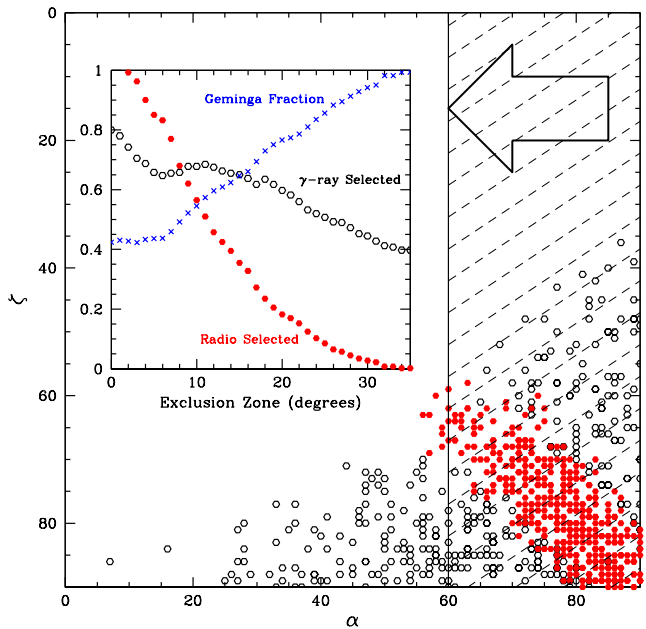}
\vskip -0.5truecm
\caption{\label{Alignment} Figure showing the effects of magnetic alignment
on the detection fractions of older ($\dot{E} \le 2.5\times10^{34}\,{\rm erg\,s^{-1}}$)
$\gamma$-ray pulsars.}
\end{figure}

	The second OG alignment effect is a reduction in $\gamma$-ray 
detections at low $\dot{E}$. In outer gap geometries, the
approach of the null charge crossing to the magnetic axis
and light cylinder occurs at higher $\dot{E}$ for aligned pulsars.
Thus gaps saturate and shut off at younger ages (i.e. higher $\dot{E}$
values) the more aligned a pulsar is.
This has pulse shape effects, but most importantly causes more
rapid depletion of the smaller $\alpha$ pulsars. In fact,
the process preferentially eliminates radio-selected 
$\gamma$-ray pulsars. We illustrate this in Figure \ref{Alignment}.
The main field in that figure shows the $\alpha-\zeta$ plane,
populated with detected $\gamma$-ray pulsars.
$\gamma$-selected pulsars are shown with open black circles, while
radio-selected pulsars are shown with filled red circles.
The radio-selected objects must have small $|\beta| = |\zeta - \alpha|$.

	As alignment proceeds, we expect that the largest $\alpha$
values will migrate to lower $\alpha$. Note that as angles
from $80^\circ - 60^\circ$ are depleted (hatched zone), many 
more radio-selected objects than $\gamma$-selected objects are lost. 
Additionally, many radio-selected objects actually become
$\gamma$-selected objects, while the converse is much more
rarely the case.
This is illustrated in the figure inset, showing the effects of an
increasingly large shift in alpha values.
The number of radio-selected detections
drops much faster than the $\gamma$-selected detections
(which in fact increases at some points, for the reason mentioned above)
and the Geminga fraction grows toward unity.

  Since the principal alignment effects amend discrepancies
seen in the data comparison, a more complete study 
seems warranted. In particular, the high sensitivity of outer 
magnetosphere models to $\alpha$ can make the effects of even 
subtle alignment more easily detected than in the radio data,
which is largely controlled by $\beta$. 
Note that lower altitude models, such as TPC, are much less 
affected by alignment.  In practice we might expect some spread 
in the alignment rates with a few objects persisting at 
large $\alpha$ to relatively late times so a detailed kinematic 
study may be required.

\section{Discussion, Conclusions, \& Predictions}
\label{Conclusion}

	We have simulated a model Milky Way Galaxy filled with
young pulsars, and by comparing the distributions of various 
observables we have been able to constrain properties of the 
true population and evaluate the relative fidelity of
several $\gamma$-ray pulsar models. To start, we found that
a radio luminosity distribution scaling with ${\dot E}$
can provide a good fit to the observations, but only if the scaling relation
is broken and plateaus at the very highest luminosities.
We also find that the $\gamma$-selected sample has so many pulsars 
with large spin-down luminosity that a short birth period distribution
($\langle P_0 \rangle \approx 50\,$ms) is required. 
These parameters also provide a good description of the
young radio pulsar population.

	In comparing with several $\gamma$-ray emission models,
all based on the vacuum dipole field structure, we find that 
Outer Gap (OG) models perform the best in almost all circumstances.
Thus, {\it for the population as a whole} there is a very statistically
significant (formally $\ge 10^7\times$) preference for the OG model -- i.e a model
dominated by radiation above the null charge surface.  However,,
there are a handful of individual objects that do {\it not}
fit easily into the predicted OG population. Such objects may
well have significant lower altitude emission, as posited by Slot Gap
(TPC-type) models. Discerning the physical parameters which select
when such emission is present will be particularly interesting.
It should also be remembered that while these vacuum calculations 
give a clear preference for OG geometries over TPC geometries, they are 
themselves not perfect fits and do not represent complete physical models.
This is evident since even the OG model does not produce $\chi^2$/DoF=1.
This is not surprising since any real magnetosphere must include some
currents and plasma affects will certainly perturb the vacuum conclusions. Thus 
it will be of interest to extend this population model/data 
comparison to the other extremum, the fully force free plasma-dominated 
SL models \citep{bssl}.

	We conclude this discussion by making future predictions of
the pulsar birthrate, detectable numbers and background contribution
for the preferred OG model.
The 6 Month sample from {\it Fermi} contains 37 $\gamma$-ray pulsars, with a
Geminga Fraction of 49\%.  At the 6 Month sensitivity level, the 
OG model produces 2194 $\gamma$-ray detections (normalized to
one pulsar birth per year) with a Geminga Fraction of 46\%. Thus the observed
pulsar count directly gives the pulsar birthrate:
$2194/37\approx59$ years per pulsar birth, or 
$1.69 \pm 0.24 ({\rm stat})\rm \ pulsars\ century^{-1}$.
This is in excellent agreement with the core collapse rate 
\citep{die06} and OB star birthrate \citep{reed05} measurements quoted in 
Section \ref{GalStruct}. It is also only $1.2\sigma$ lower than the SN rate
computed by \citet{li10}, which has the smallest formal error bars.

	It is interesting to compare this birthrate estimate
to the independent O-B2 star birthrate estimate made in this work,
based on the local massive star density ($2.4\rm \ OB\ century^{-1}$).
The first thing to note is that our $\gamma$-ray pulsar birthrate
is a large fraction of these progenitor rates. For example, at
71\% of the O-B2 star rate, a large fraction of these stars {\it must}
produce $\gamma$-ray pulsars. Indeed we check if culling the lowest
mass class (B2, $\sim 9 M_\odot$) is acceptable.
With this cut the local massive star birthrate drops from  
$40\rm \ OB\ kpc^2\ Myr^{-1}$ to $21.5\rm \ OB\ kpc^2\ Myr^{-1}$,
which corresponds to a Galactic rate of $1.3\rm \ OB\ century^{-1}$.
This is only 75\% of the rate needed to produce our inferred
$\gamma$-ray pulsar population. This is also 2.1$\sigma$ lower than
the \citet{li10} Galactic supernova rate. Thus we conclude that
B2 stars should produce SNe and contribute to the young
pulsar population.

	Perhaps even more interesting is the direct comparison
between our pulsar birthrate and the core collapse rate. The observed
$\gamma$-ray pulsars contribute $>75\%$ of the (relatively high)
inferred SN rate in this study. Thus we conclude that the
$\gamma$-ray pulsar sample must be an excellent census of the
local core collapse products. These rates imply that no
more than 25\% of the core collapses can produce slow
$P_0$ injected pulsars, RRATS, Central Compact Sources
in Supernova Remnants, magnetars and other exotica. 
Similar conclusions have already been reached by \citet{kk08}
using the radio surveys.  Thus our new population estimate,
derived from the {\it Fermi} sample and sensitive to $\gamma$-ray
rather than radio beaming effects, supports the picture that 
energetic radio pulsars dominate the neutron star birthrate, and that
more exotic objects cannot contribute a dominant fraction of the
neutron star population unless they represent a later phase in
the evolution of energetic, $\gamma$-ray producing pulsars.

	In outer magnetosphere models there are, of course, some 
pulsars whose bright $\gamma$-ray beam misses the Earth while
the radio beam does not.  These objects can be detected in the radio
but will be absent in the $\gamma$ rays, despite the fact that they
are $\gamma$-ray active.  This especially occurs for geometries with small
$\alpha$ and small $\zeta$ (see Figure \ref{Alignment} where radio-only pulsars
continue to the upper left); for a few objects, known angles place
them in this region, e.g PSR J1930+1852 \citep{nr08,psrcat}. Of course,
fainter low altitude emission aligned with the radio could still
be visible; PSR J0659+1414 could be such a case. Our population
estimates suggest that there should be a modest number of
objects in this radio loud, $\gamma$-ray missed category.
At the sensitivity of the 1 year sample we would expect 
$\sim 10$ radio-detected, $\gamma$-ray undetected pulsars
with ${\dot E} >10^{33.5}$ and
${\dot E}^{1/2}/d^2$ as large as that of the faintest detected
$\gamma$-ray pulsars; again, these are objects not seen in the $\gamma$-rays
simply due to beaming effects.
Most of these have ${\dot E} < 10^{35}\,{\rm erg\, s^{-1}}$. As the
$\gamma$-ray sensitivity grows, the fraction of such objects decreases.
However there are also pulsars whose radio and $\gamma$-ray beams
both miss the Earth.  At the 1-year sensitivity nearly half of
the pulsars with ${\dot E}^{1/2}/d^2$ above that of the faintest detections
are such 'Isolated Neutron Stars' (INS). 
Such radio-only pulsars and INS are implicitly included in the 
birthrate estimates above.

	Given the successes of the OG model at matching the observed
population, it is natural to ask what this scenario predicts 
as we acquire additional observations.  For the birthrate above,
we can lower the $\gamma$-ray threshold and make predictions of the 
number of detections as a function of {\it Fermi} observation time.
Our ``Current'' {\it Fermi} sample is based on approximately
one full year of data; extrapolating from the 6-month normalization, 
we find that the model predicts 23 $\gamma$-selected pulsars and 
26 radio-selected pulsars.  The actual observed sample contains 
24 $\gamma$-selected pulsars and 26 radio-selected pulsars,
an excellent match to the observations. Similarly 
after five years exposure the model predicts 49 and 44 detections
and after ten years 65 and 55, $\gamma$-selected and radio-selected
detections, respectively. With the improved 10-year $\gamma$-ray sensitivity,
the number of radio-detected, $\gamma$-ray undetected pulsars
grows at a similar rate; we expect $\sim 27$ such objects 
with $\dot E^{1/2}/d^2$ large enough that we would expect detection.
The number of INS whose $\gamma$-ray beam 
would be detectable, if directed at Earth, grows to 
$\sim 150$ objects, still about 1/2 of the total population. Most
are again low ${\dot E}$ pulsars whose relatively narrow radio and $\gamma$-ray
beams only sweep a small fraction of the sky.

	Note that we can hope that the {\it actual} $\gamma$-ray
pulsar numbers will be larger by a modest fraction as analysis and search techniques
improve and as deep radio observations lower the effective survey threshold.
Note also that these numbers do not include the very substantial
population of recycled pulsars now being detected; the
equivalent analysis of this population will be prosecuted in future work.

	The trend towards higher Geminga Fraction above in these future
predictions is due to two effects. First, as {\it Fermi} probes 
larger distance scales our radio pulsar sample will not be as complete.
In practice, as just noted, deeper radio searches will
mitigate this trend. Of course, we still expect a substantial 
number of very low flux density sources will still remain non-detected,
since the present very deep radio searches on many LAT sources
indicates that these are a substantial part of the Geminga population.
The second effect increasing the Geminga Fraction is
a change in the $\dot{E}$ distribution of the detected pulsars.
The five and ten year model samples have a higher proportion of mid to low
luminosity objects ($\dot{E}\lesssim 10^{35}\,{\rm erg\ s}^{-1}$),
and thus proportionately fewer energetic pulsars. With longer periods
at lower $\dot{E}$ come narrower radio beams and fewer detections.
If alignment, as discussed above, is included in the models this 
contribution to the Geminga fraction will be even larger.

	The final number to report is the $\gamma$-ray background 
supplied by unresolved pulsars. To compute this we simply
sum up the emission from sources not individually detected
(at a given LAT exposure time). To estimate the spectral
content of this contribution, we assign each modeled pulsar
a power-law spectral index $\Gamma$ and exponential cut-off energy
$E_c$.  The assigned value is inferred from the $\Gamma({\dot E})$ 
and $E_c(B_{LC})$ trends visible in Figures 7 and 8
of \citet{psrcat} and approximated here by
\begin{eqnarray}
\rm{E_c}=[-0.45 + 0.71\,\rm{log(B_{LC})}]\,{\rm GeV} \nonumber \\
\Gamma=-4.10 + 0.156\,\rm{log(\dot{E})} .
\end{eqnarray}
These model spectra can be used to sum up the effective young
pulsar background spectrum. In practice we compute 0.1-1\,GeV
and 1-10\,GeV sky maps of this emission to look for any
interesting spatial distribution. Some evidence of cluster structure
is seen in these maps, but in any event the contribution to the
total Galactic background is small. 

	At the 6 month sensitivity limit we find that pulsars 
supply $4.4\times 10^{-9}\, \rm erg\ cm^{-2}\ s^{-1}$ (1.8\%) of 
the total background flux in the 0.1-1 GeV band and
$3.6 \times 10^{-9}\, \rm erg\ cm^{-2}\ s^{-1}$ (2.8\%) of the flux in
the 1-10 GeV band (integrated over the full sky and compared against the
{\it Fermi} Collaboration's publicly available background skymap, gll\_iem\_v02).
Of course, as time progresses and more pulsars are individually detected,
the total unresolved flux from background pulsars goes down.
After 10 years of observation, approximately half of the total flux
unresolved at 6 months will have been detected as
individual sources.  
We would then find that pulsars contribute only 1.0\% and 1.5\% of the 0.1-1 Gev
and 1-10 GeV background bands, respectively. We remind the reader
that this does not include the contribution of recycled pulsars.
As noted, some studies \citep{fgl10} estimate that this may be quite substantial
at intermediate latitude. A treatment with improved luminosity, beaming
and evolutionary effects, as presented here for the young
pulsar population, seems very desirable.

\acknowledgements

This work was supported in part by NASA grants NAS5-00147 and NNX10AD11G. 
K.P.W. was supported by NASA under contract NAS5-00147.
This work made extensive use of the ATNF pulsar catalog at
http://www.atnf.csiro.au/research/pulsar/psrcat/
\citep{m05}.

\end{document}